\documentclass{webofc}
\usepackage[varg]{txfonts}   
\begin{document}
\title{Three-body radiative capture reactions}
%
%

\author{\firstname{J.} \lastname{Casal}\inst{1,2}\fnsep\thanks{\email{jcasal@us.es}} \and
        \firstname{M.} \lastname{Rodríguez-Gallardo}\inst{2}\fnsep \and
        \firstname{J. M.} \lastname{Arias}\inst{2}\fnsep \and
        \firstname{J.} \lastname{Gómez-Camacho}\inst{2,3}\fnsep
}

\institute{European Centre for Theoretical Studies in Nuclear Physics and Related Areas (ECT$^*$) and Fondazione Bruno Kessler, Villa Tambosi, Strada delle Tabarelle 286, I-38123 Villazzano (TN), Italy
\and
           Departamento de F\'{\i}sica At\'omica, Molecular y Nuclear, Facultad de F\'{\i}sica, Universidad de Sevilla, Apartado 1065, E-41080 Sevilla, Spain
\and
           Centro Nacional de Aceleradores, Universidad de Sevilla, Junta de Andalucía, CSIC, Tomas Alva Edison 7, E-41092 Sevilla, Spain
          }

\abstract{%
   Radiative capture reaction rates for $^6$He, $^9$Be and $^{17}$Ne formation at astrophysical conditions are studied within a three-body model using the analytical transformed harmonic oscillator method to calculate their states. An alternative procedure to estimate these rates from experimental data on low-energy breakup is also discussed.
}
\maketitle
\section{Introduction}
\label{intro}
Radiative capture reactions are crucial for the stellar models aiming to describe the evolution in composition, energy production and temperature structure of different astrophysical environments \cite{Langanke01}. For two-body reactions involving stable nuclei, the direct experimental measurement of the relevant cross section gives directly the astrophysical quantity. In many cases, however, reaction cross sections may not be measured directly. This may occur if the initial nucleus is exotic \cite{Arnould99}, or when the capture process is a three-body reaction \cite{Sumiyoshi02}.
Three-body radiative capture processes are essential in overcoming the $A=5,8$ instability gaps, and traditionally they have been described as two-step sequential reactions \cite{Angulo99}. However, it has been shown that the direct three-body capture may play an important role at low temperatures, where the particles have no energy to populate intermediate resonances \cite{Garrido11}. Radiative capture reaction rates can be obtained from the inverse photodissociation cross section, 
which can be calculated within a proper three-body model for the compound nucleus \cite{RdDiego10}. In this work, we summarize some recent results on three-body capture reactions and discuss future developments.

\section{Theoretical framework and applications}
\label{sec-1}

The energy-averaged reaction rate for three-body \mbox{$a+b+c \rightarrow$ $A + \gamma$} radiative capture processes can be calculated from the inverse photodissociation reaction as a function of the temperature  $T$ by the expression~\cite{RdDiego10,JCasal13}
\begin{equation}
\langle R_3(\varepsilon)\rangle(T) =  \frac{\mathcal{C}_3}{\left(k_B T\right)^{3}}  e^{\frac{|\varepsilon_B|}{k_B T}} \int_{|\varepsilon_B|}^\infty d\varepsilon_\gamma~\varepsilon_\gamma^2 \sigma_\gamma(\varepsilon_\gamma) e^{\frac{-\varepsilon_\gamma}{k_B T}},
\label{eq:aRE}
\end{equation}
where $\varepsilon=\varepsilon_\gamma+\varepsilon_B$ is the initial three-body kinetic energy, $\varepsilon_\gamma$ is the energy of the photon emitted, and $\varepsilon_B$ is the ground-state energy of the compound nucleus $A$. Here, $\mathcal{C}_3$ is a constant depending on the masses and spins of the initial particles. The photodissociation cross-section $\sigma_\gamma(\varepsilon_\gamma)$ of the nucleus $A$ can be expanded into electric and magnetic multipoles, $\lambda$, as~\cite{Forseen03}
\begin{equation}
\sigma_\gamma^{(\lambda)}(\varepsilon_\gamma)=\frac{(2\pi)^3 (\lambda+1)}{\lambda[(2\lambda+1)!!]^2}\left(\frac{\varepsilon_\gamma}{\hbar c}\right)^{2\lambda-1}\frac{dB({\cal O}\lambda)}{d\varepsilon},
\label{eq:xsection}
\end{equation}
where $B({\cal O}\lambda)$ is the order $\lambda$ of electric or magnetic transition probability (${\cal O} = E, M$). The previous expressions make no assumption about the reaction mechanism, thus including sequential and direct, resonant and non-resonant contributions on an equal footing.
In computing the reaction rate given by Eq.~(\ref{eq:aRE}), only the corresponding transition probability distributions of the compound nucleus $A$ are required. To this purpose, we recently applied the analytical transformed harmonic oscillator (THO) method to study the Borromean nuclei $^6$He \cite{JCasal13}, $^9$Be \cite{JCasal14,JCasal15} and $^{17}$Ne \cite{JCasal16} and their formation at astrophysical conditions. 

\subsection{Formation of $\boldsymbol{^6\text{He},{ ^9\text{Be}}}$ and $\boldsymbol{^{17}\text{Ne}}$} 
\label{sec-2.1}
\begin{figure*}[b]
	\centering
	\includegraphics[width=0.39\linewidth]{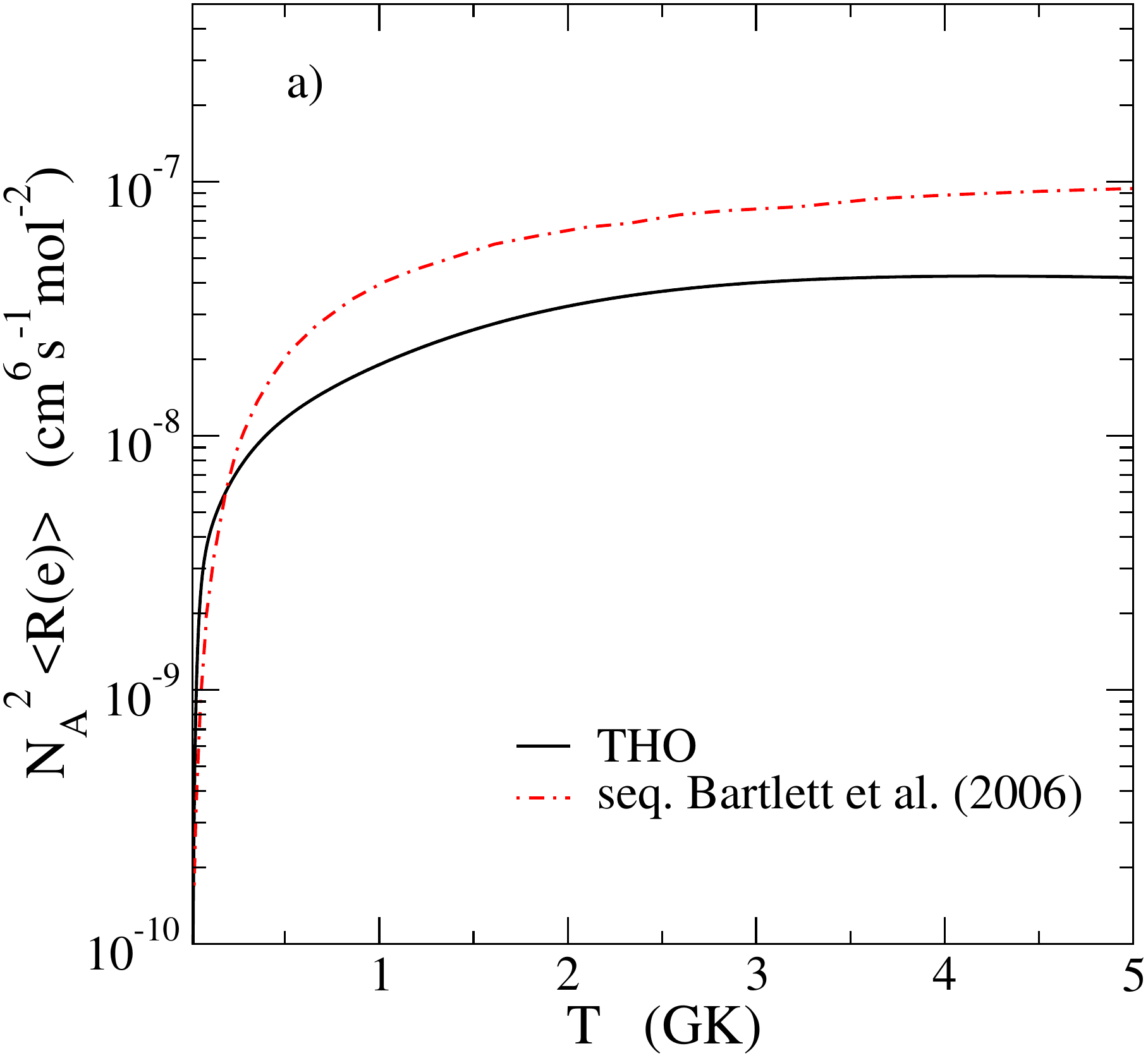}\hspace{15pt}\includegraphics[width=0.395\linewidth]{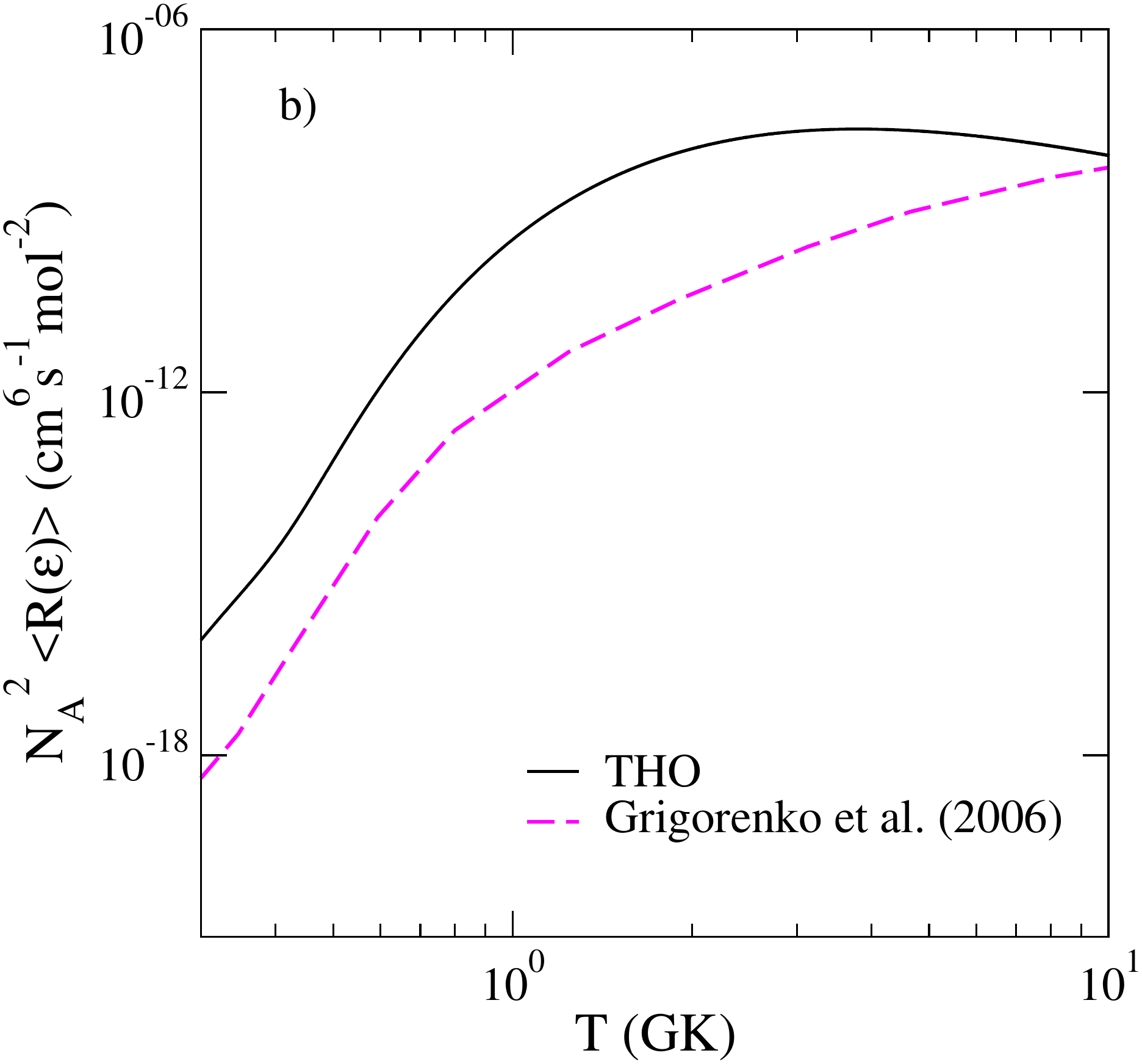}
	\caption{Reaction rate of (a) $\alpha(2n,\gamma){^6\text{He}}$ and (b) $^{15}\text{O}(2p,\gamma){^{17}\text{Ne}}$ using the THO method to generate $^6$He and $^{17}$Ne states. Our results are compared with previous estimations from Refs.~\cite{Bartlett06,Grigorenko06}.}
	\label{fig:rate1}       
\end{figure*}

It has been proposed that the $\alpha(2n,\gamma){^6\text{He}}$ reaction may play a role in the evolution of neutron star mergers \cite{Bartlett06}, one of the possible scenarios for the r-process nucleosynthesis. 
The reaction rate of the $\alpha(2n,\gamma){^6\text{He}}$ process is governed by $E1$ transitions between the $0^+$ ground state and $1^-$ continuum states \cite{RdDiego10}. In Fig.~\ref{fig:rate1}a, we show our computed reaction rate for $^6$He formation, using the analytical THO method to generate $^6$He states. 
The details about these structure calculations can be found in Ref.~\cite{JCasal13}. We compare our results with a previous sequential estimation assuming a dineutron formation \cite{Bartlett06}, showing important differences between them. 

The two-proton capture reaction $^{15}\text{O}(2p,\gamma){^{17}\text{Ne}}$ may play a role for the rp-process in Type I X-ray bursts \cite{Gorres95}. As in the preceding example, its formation is governed by $E1$ transitions. In this case, it is due to the presence of a low-energy 1/2$^+$ resonance in $^{17}$Ne. 
Details can be found in Ref.~\cite{JCasal16}. 
Our computed reaction rate, shown in Fig.~\ref{fig:rate1}b, is several orders of magnitude larger than the previous calculations in Ref.~\cite{Grigorenko06}. The possible implications of this difference should be investigated.

\begin{figure*}[t]
	\centering
	\includegraphics[width=0.44\linewidth]{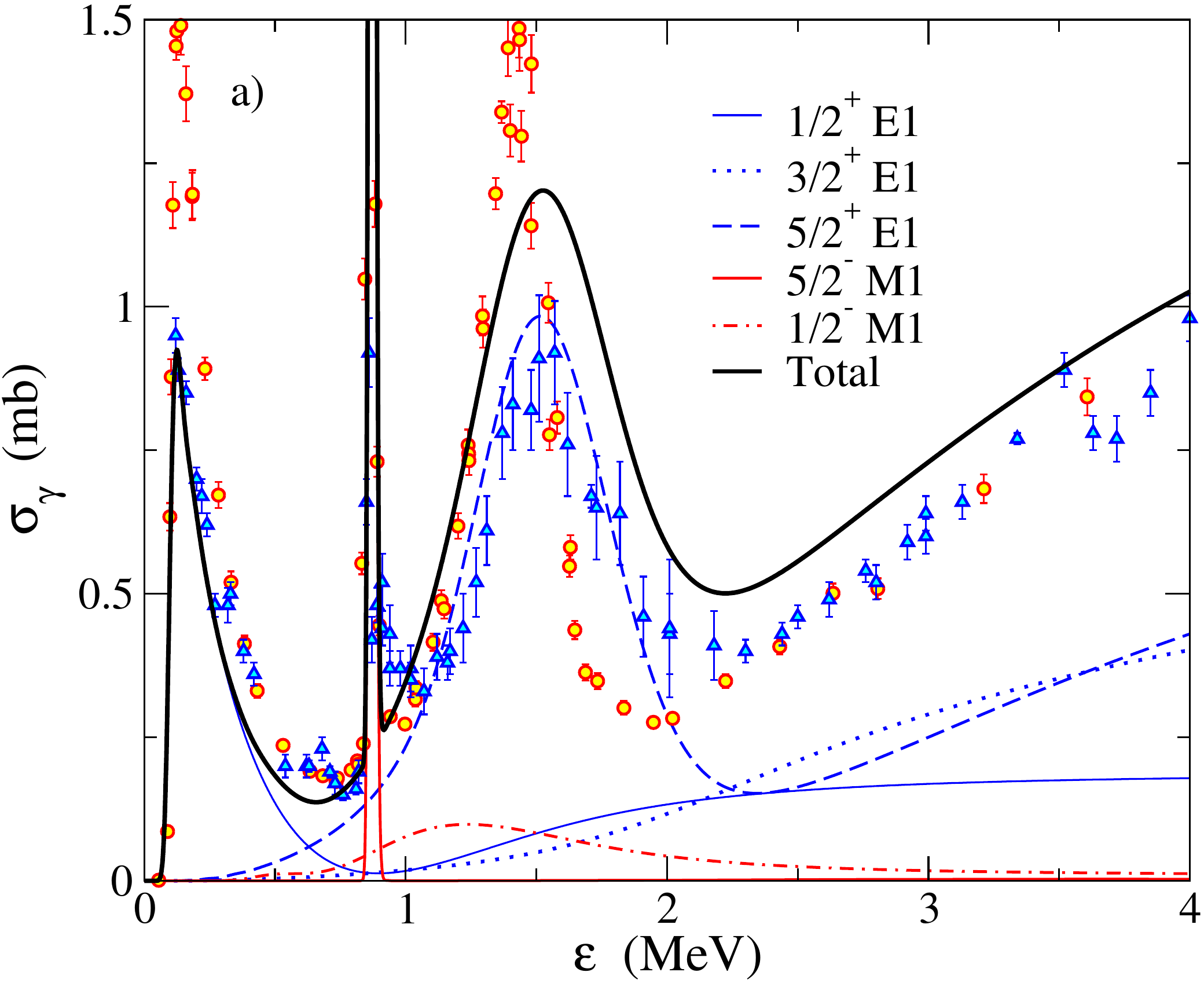}\hspace{15pt}\includegraphics[width=0.39\linewidth]{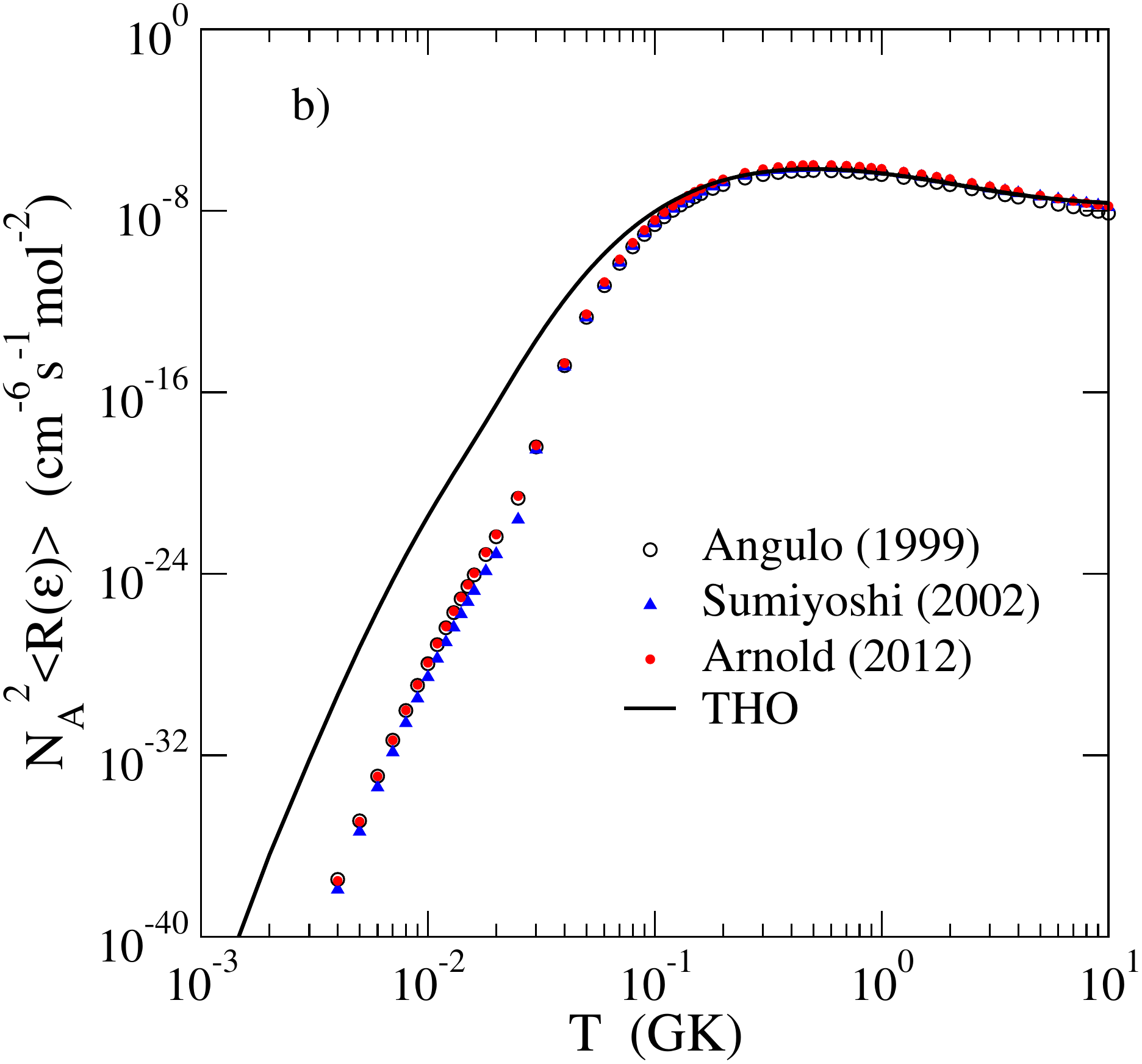}
	\caption{(a) Photodissociation cross section of $^9$Be using the THO method compared with the data from Refs. \cite{Sumiyoshi02} (triangles) and \cite{Arnold12} (circles). (b) Reaction rate of $\alpha(\alpha n,\gamma){^9\text{Be}}$ compared with sequential estimations from Refs. \cite{Angulo99,Sumiyoshi02,Arnold12} . 
}
	\label{fig:9Be}       
\end{figure*}

The formation of $^9$Be has been also linked to the r-process, in this case in type II supernovae~\cite{Sumiyoshi02,Arnold12}. The reaction rate of $\alpha(\alpha n,\gamma){^9\text{Be}}$ is governed by low-energy resonances in $^9$Be \cite{RdDiego10}, especially the near-threshold 1/2$^+$ state. 
In Fig.~\ref{fig:9Be}, we compare our calculations with the experimental data on the photodissociation cross section for $^9$Be, including $E1$ and $M1$ transitions, and we show the corresponding reaction rates. Details can be found in Ref.~\cite{JCasal14}. Our calculations agree with sequential estimations at high temperature, but show an enhancement at low temperature coming from the three-body direct $E1$ capture to low-energy 1/2$^+$ states. 
This effect cannot be described using the definition of the rate in Refs.~\cite{Sumiyoshi02,Arnold12}, where a sequential process is assumed.

\section{Reaction rates from inclusive breakup data}
\label{sec-3}

\begin{figure}[h]
	\vspace{-15pt}
	\centering
	\hspace{-1cm}\includegraphics[width=0.72\linewidth]{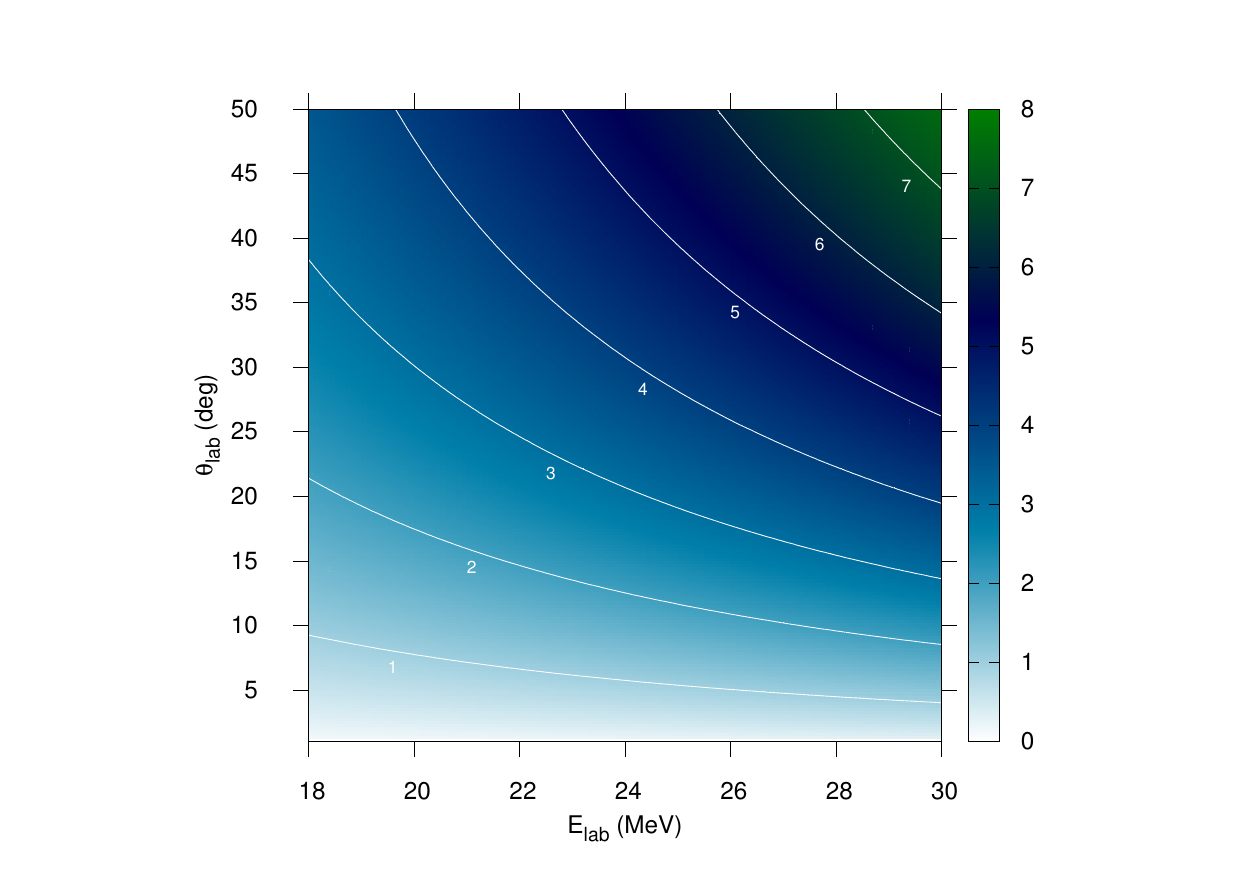}\hspace{-1cm}\vspace{-10pt}
	\caption{Temperatures, in a color scale, at which the reaction rates for the $\alpha(2n,\gamma){^6\text{He}}$ process can be obtained, as a function of the scattering angle and bombarding energy in the $^6$He $+$ $^{208}$Pb reaction. This is obtained from the definition of the collision time and the relation $t=1/k_BT$. White lines represent fixed temperatures in GK.
}
	\label{fig:RP}
    \vspace{-20pt}
\end{figure}

Radiative capture rates computed within different theoretical frameworks may exhibit important differences, as shown in Fig.~\ref{fig:rate1}. We recently proposed an alternative method to estimate these reaction rates from inclusive low-energy breakup data on the compound nucleus \cite{JCasal16R}. When the process is $E1$-dominated, such as in the previous examples, the reaction rate can be obtained as

\begin{equation}
\langle R_3(\varepsilon)\rangle(T) = C_3' t^{3}e^{|\varepsilon_B|t} \frac{d^2}{dt^2}\left(\frac{1}{t^2}P_r(t)\right),
\label{eq:RP}
\end{equation}
where $P_r$ is the reduced breakup probability~\cite{JPFerGar13} and $t=\frac{a_0}{\hbar v}\left(\pi+\frac{2}{\sin(\theta/2)}\right)$ is the collision time, associated with the temperature through $t=1/k_BT$. Here, $a_0$ is half the classical closest approach distance, and $v$ is the relative projectile-target velocity. Equation~(\ref{eq:RP}) is valid only at small scattering angles, where the process is Coulomb dominated, and allows us to explore different astrophysical temperatures by measuring the breakup at different kinematic conditions. This is shown in Fig. \ref{fig:RP} for $^6$He $+$ $^{208}$Pb, which can provide information on the $\alpha(2n,\gamma){^6\text{He}}$ reaction. In Ref.~\cite{JCasal16R}, the method was applied to the case of $^{11}$Li, for which suitable experimental data were available~\cite{JPFerGar13}. For $^6$He, however, the existing data do not cover an optimal angular region for the method to be valid. Therefore, a dedicated experiment is required to make a reliable estimation of the rate and to assess the validity of the different theoretical approaches. Such data have to be restricted to the low-energy and small-angle sector of Fig.~\ref{fig:RP}, providing the reaction rates in the region between 1-3 GK.

\section*{Acknowledgements}

This work has been partially supported by the Spanish Ministerio de Economía y Competitividad under Project No.~FPA2016-77689-C2-1-R, No.~FIS2014-53448-c2-1-P, and No.~FIS2014-51941-P, by Junta de Andalucía under Group No.~FQM-160 and Project No.~P11-FQM-7632, and by the European Union's Horizon 2020 research and innovation program under grant agreement No.~654002. M.~R.-G.~acknowledges postdoctoral support from the Universidad de Sevilla under the V Plan Propio de Investigación, Contract No.~USE-11206-M.

%
\bibliography{bibfile}

\begin{thebibliography}{17}

\bibitem{Langanke01}
K.~Langanke, M.~Wiescher, Rep. Prog. Phys. \textbf{64}, 1657 (2001)

\bibitem{Arnould99}
M.~Arnould, K.~Takahashi, Rep. Prog. Phys. \textbf{62}, 395 (1999)

\bibitem{Sumiyoshi02}
K.~Sumiyoshi, H.~Utsunomiya, S.~Goko, T.~Kajino, Nucl. Phys. A \textbf{709},
  467 (2002)

\bibitem{Angulo99}
C.~{Angulo {\it et al.}}, Nucl. Phys. A \textbf{656}, 3 (1999)

\bibitem{Garrido11}
E.~Garrido, R.~de~Diego, D.V. Fedorov, A.S. Jensen, Eur. Phys. J. A
  \textbf{47}, 102 (2011)

\bibitem{RdDiego10}
R.~de~Diego, E.~Garrido, D.V. Fedorov, A.S. Jensen, Eur. Phys. Lett.
  \textbf{90}, 52001 (2010)

\bibitem{JCasal13}
J.~Casal, M.~Rodr{\'i}guez-Gallardo, J.M. Arias, Phys. Rev. C \textbf{88},
  014327 (2013)

\bibitem{Forseen03}
C.~Forss{\'e}n, N.B. Shul'gina, M.V. Zhukov, Phys. Rev. C \textbf{67}, 045801
  (2003)

\bibitem{JCasal14}
J.~Casal, M.~Rodr{\'i}guez-Gallardo, J.M. Arias, I.J. Thompson, Phys. Rev. C
  \textbf{90}, 044304 (2014)

\bibitem{JCasal15}
J.~Casal, M.~Rodr\'{\i}guez-Gallardo, J.M. Arias, Phys. Rev. C \textbf{92},
  054611 (2015)

\bibitem{JCasal16}
J.~Casal et~al., Phys. Rev. C \textbf{94}, 054622 (2016)

\bibitem{Bartlett06}
A.~Bartlett et~al., Phys. Rev. C \textbf{74}, 015802 (2006)

\bibitem{Gorres95}
J.~G\"orres, M.~Wiescher, F.~Thielemann, Phys. Rev. C \textbf{51}, 392 (1995)

\bibitem{Grigorenko06}
L.V. Grigorenko, K.~Langanke, N.~Shul'gina, M.~Zhukov, Phys. Lett. B
  \textbf{641}, 254 (2006)

\bibitem{Arnold12}
C.W. Arnold et~al., Phys. Rev. C \textbf{85}, 044605 (2012)

\bibitem{JCasal16R}
J.~Casal et~al., Phys. Rev. C \textbf{93}, 041602(R) (2016)

\bibitem{JPFerGar13}
J.P. {Fernández-García {\sl et al.}}, Phys. Rev. Lett. \textbf{110}, 142701
  (2013)

\end{thebibliography}
%
%
%
%

\end{document}